%% file: main.tex
\documentclass[10pt,twocolumn,english,aps,pra,superscriptaddress,amsmath,amssymb,floatfix,footinbib]{revtex4-2}
\usepackage{times}
\usepackage[T1]{fontenc}
\usepackage[utf8]{inputenc}
\usepackage[english]{babel}
\addto\captionsenglish{}
\usepackage{amsmath,amsfonts,amssymb}
\usepackage{stmaryrd}
\usepackage{graphicx}
\usepackage{dcolumn,multirow,bm}
\usepackage{chemformula}
\usepackage[version=4]{mhchem}
\usepackage{siunitx}
\usepackage{textcomp}
\usepackage{xurl}
\usepackage[colorlinks=true,citecolor=blue,linkcolor=magenta]{hyperref}
\hypersetup{pdftitle={Unlocking scalable manufacturing of high-performance integrated electro-optic circuits via substrate engineering},pdfauthor={Jiachen Cai et al.}}
\DeclareSIUnit \dbc {dBc}
\newcommand{\LT}{LiTaO$_3$}
\newcommand{\LN}{LiNbO$_3$}
\newcommand{\silica}{SiO$_2$}

\newcommand{\um}{\text{\textmu m}}


\begin{document}
\title{Unlocking scalable manufacturing of high-performance integrated electro-optic circuits via substrate engineering}
\author{Jiachen~Cai}
\thanks{These authors contributed equally to this work.}
\affiliation{State Key Laboratory of Materials for Integrated Circuits, Shanghai Institute of Microsystem and Information Technology, Chinese Academy of Sciences, Shanghai, China}
\affiliation{Institute of Physics, Swiss Federal Institute of Technology Lausanne (EPFL), CH-1015 Lausanne, Switzerland}
\author{Junyin~Zhang}
\thanks{These authors contributed equally to this work.}
\affiliation{Institute of Physics, Swiss Federal Institute of Technology Lausanne (EPFL), CH-1015 Lausanne, Switzerland}
\author{Marcus~Albrechtsen}
\thanks{These authors contributed equally to this work.}
\affiliation{Institute of Physics, Swiss Federal Institute of Technology Lausanne (EPFL), CH-1015 Lausanne, Switzerland}
\author{Shuhang~Zheng}
\affiliation{Institute of Physics, Swiss Federal Institute of Technology Lausanne (EPFL), CH-1015 Lausanne, Switzerland}
\author{Chengli~Wang}
\affiliation{State Key Laboratory of Materials for Integrated Circuits, Shanghai Institute of Microsystem and Information Technology, Chinese Academy of Sciences, Shanghai, China}
\author{Mengxin~Lin}
\affiliation{Institute of Physics, Swiss Federal Institute of Technology Lausanne (EPFL), CH-1015 Lausanne, Switzerland}
\author{Xinru~Ji}
\affiliation{Institute of Physics, Swiss Federal Institute of Technology Lausanne (EPFL), CH-1015 Lausanne, Switzerland}
\author{Yichi~Zhang}
\affiliation{Institute of Physics, Swiss Federal Institute of Technology Lausanne (EPFL), CH-1015 Lausanne, Switzerland}
\author{Zheru~Qiu}
\affiliation{Institute of Physics, Swiss Federal Institute of Technology Lausanne (EPFL), CH-1015 Lausanne, Switzerland}
\author{Xin~Ou}
\email{ouxin@mail.sim.ac.cn}
\affiliation{State Key Laboratory of Materials for Integrated Circuits, Shanghai Institute of Microsystem and Information Technology, Chinese Academy of Sciences, Shanghai, China}
\author{Tobias J. Kippenberg}
\email{tobias.kippenberg@epfl.ch}
\affiliation{Institute of Physics, Swiss Federal Institute of Technology Lausanne (EPFL), CH-1015 Lausanne, Switzerland}
\affiliation{Institute of Electrical and Micro Engineering, Swiss Federal Institute of Technology Lausanne (EPFL), CH-1015 Lausanne, Switzerland}
\maketitle

\noindent\textbf{Thin-film ferroelectrics such as lithium niobate, lithium tantalate and barium titanate support high-density photonic integrated circuits with strong Pockels nonlinearity, enabling a new generation of ultrafast electro-optic applications ranging from classical optical communications to quantum transduction.
However, high-volume manufacturing of these platforms in standard silicon foundries, essential for their widespread deployment, remains challenging.
We identify a previously unrecognized limitation to room-temperature ferroelectric electro-optics---microwave loss exacerbated by plasma-induced parasitic surface conduction during foundry-compatible processing.
Adopting techniques from radio-frequency electronics, we overcome this bottleneck with a photonic trap-rich lithium-tantalate-on-insulator platform combining carrier-trapping silicon with a hydrogen-free buried oxide.
This yields ultralow optical losses, broader modulator bandwidths and lower electro-optic comb drive power.
The microwave-loss advantage persists below the superconducting transition of the aluminum electrodes into the millikelvin regime.
These results enable scalable, process-robust ferroelectric electro-optics within the silicon manufacturing ecosystem.}

\subsection{Introduction}
Efficient conversion between electrical and optical signals is the cornerstone of high-speed optical interconnects, alleviating the growing bandwidth and energy bottlenecks in data-center networking~\cite{zhou2024silicon} and enabling compact, energy-efficient co-integrated photonic--electronic systems~\cite{daudlin2025three}.
    The photonic generation and processing of radio-frequency (RF) signals over broad instantaneous bandwidths also form the basis of integrated microwave-photonic systems~\cite{marpaung2019integrated}.
   Thin-film lithium niobate (\LN) and lithium tantalate (\LT) have emerged as platforms for low-loss, high-density photonic integrated circuits (PICs), owing to their strong Pockels response, broad optical transparency and the availability of mature on-insulator substrates enabled by material-efficient ion slicing~\cite{boes2023lithium, wang2024lithium}.
    Lithium tantalate further benefits from low birefringence, reduced photorefractive effects and an established high-volume wafer supply chain serving the RF filter industry~\cite{wang2024lithium}.
    These platforms have enabled high-speed, low-voltage modulators~\cite{wang2018integrated}, electro-optic (EO) frequency comb generation~\cite{zhang2019broadband} and ultrafast tunable lasers~\cite{siddharth2025ultrafast} for applications ranging from telecommunications to signal processing~\cite{hu2025integrated}.
    Recent room-temperature demonstrations have extended their EO capabilities to millimeter-wave and terahertz frequencies~\cite{zhang2026integrated,lampert2025photonics}.
    At cryogenic temperatures, these platforms are also the basis for quantum coherent transduction between microwave and optical frequencies ~\cite{warner2025coherent}.
    To achieve low-voltage, power-efficient and broadband EO operation in traveling-wave optical modulators, the microwave drive must propagate with low attenuation over millimeter-to-centimeter interaction lengths while remaining velocity matched to the optical wave~\cite{shen2024photonic,wang2024ultrabroadband}.
    These microwave constraints become particularly stringent in quantum interconnect schemes using resonant EO transduction, where optical loss, microwave dissipation and phase-matched coupling govern the efficiency of microwave-to-optical frequency conversion~\cite{holzgrafe2020cavity}.

    Future adoption of these next-generation ferroelectric PICs at scale will require wafer-scale, foundry-compatible fabrication that preserves device performance.
    Here, we report that plasma-assisted processing can introduce excess microwave loss, severely degrading the room-temperature EO performance of devices fabricated on commercial thin-film \LT-on-insulator (LTOI) wafers.
    Compared with chip-scale laboratory fabrication, volume processing of these etch-resistant ferroelectric materials typically involves deep-ultraviolet (DUV) lithography and extensive plasma-assisted processing, including dry etching and dense dielectric encapsulation.
    During such processing, plasmas based on gases including argon, oxygen and silane give rise to vacuum-ultraviolet (VUV) radiation at wavelengths below approximately 200~nm, which can induce additional positive oxide-trapped charges in SiO$_2$~\cite{dimaria1977location,espinho2017vacuum}.
    Although high-resistivity silicon (HRSi) handle wafers compatible with complementary metal--oxide--semiconductor processing are employed to limit bulk substrate conduction, the accumulated oxide charges can induce parasitic surface conduction (PSC) at the oxide--silicon interface.
    This mechanism creates an additional microwave loss channel, which has long been recognized in RF electronics~\cite{wu1999sio,lederer2005new,lederer2005effective} but has so far received little attention within the ferroelectric photonics community.
    The observed EO performance degradation highlights a barrier to translating laboratory performance into scalable ferroelectric electro-optics, as well as revealing a previously unreported device-level manifestation of PSC in room-temperature EO operation.

    Trap-rich substrates, originally developed for commercial RF silicon-on-insulator (RF-SOI) electronics~\cite{raskin2013soitec, ali2014rf, esfeh2017rf}, provide an established solution to suppressing this interfacial conduction.
    Amorphous silicon (a-Si) or polycrystalline silicon (poly-Si) layers at the \silica--Si interface capture free carriers and preserve the effective substrate resistivity.
    However, adapting this strategy to ferroelectric photonics imposes requirements beyond those of conventional RF-SOI substrates.
    In RF electronics, the buried oxide mainly serves as an electrical isolation layer, and plasma-enhanced chemical vapor deposition (PECVD) with hydrogen-containing precursors (e.g., SiH$_4$ and Si(OC$_2$H$_5$)$_4$ (TEOS)) can often provide sufficient RF performance~\cite{esfeh2018rf, vandermolen2021characterization}.
    By contrast, the thick buried oxide is part of the optical stack in PICs, where hydrogen- and hydroxyl-rich oxides can introduce optical absorption~\cite{ay2004comparative}.
    Although a poly-Si layer with a PECVD buried oxide has recently enabled operation at frequencies of several gigahertz in \LN-on-insulator (LNOI) microwave resonators at cryogenic temperatures~\cite{shen2024parasitic}, this RF-SOI-like implementation leaves open the challenge of combining carrier trapping with low-optical-loss buried oxides and truly unlocking broadband, low-loss EO functionality on thin-film ferroelectric platforms.
    Meeting these optical requirements using hydrogen-containing deposition typically involves extended annealing above 1000$^\circ\mathrm{C}$ to remove hydrogen-related species~\cite{yongheng2006study,jin2020deuterated}.
   Yet such thermal processing also anneals the interfacial defect states responsible for carrier trapping~\cite{vandermolen2025traps}.

    We overcome this substrate-level challenge by engineering a 100-mm photonic trap-rich LTOI platform that simultaneously exhibits low microwave attenuation and low optical propagation loss.
    We benchmark three classes of room-temperature devices against normal-LTOI controls fabricated under the same process conditions.
    Reference coplanar waveguides (CPWs) on the trap-rich substrate maintain microwave attenuation below 1.6~dB/cm up to 70 GHz.
    The 7-mm-long traveling-wave Mach--Zehnder modulators (MZMs) extend the 3-dB EO bandwidth beyond 67 GHz.
    Resonant EO comb devices produce a spectrum spanning more than 150 nm and require 76.5\% less on-chip RF power to reach the same fitted comb-envelope slope.
    Beyond room-temperature operation, measurements of geometry-matched aluminum (Al) T-resonators show that this microwave-loss advantage persists below the superconducting transition of the electrodes into the millikelvin regime.
    Our method can be directly transferable to other material platforms and these results thus establish photonic trap-rich substrate engineering as a route to preserving broadband EO performance while retaining compatibility with high-volume silicon foundry manufacturing.

	\begin{figure*}[t!]
\centering
\includegraphics[width=0.95\textwidth]{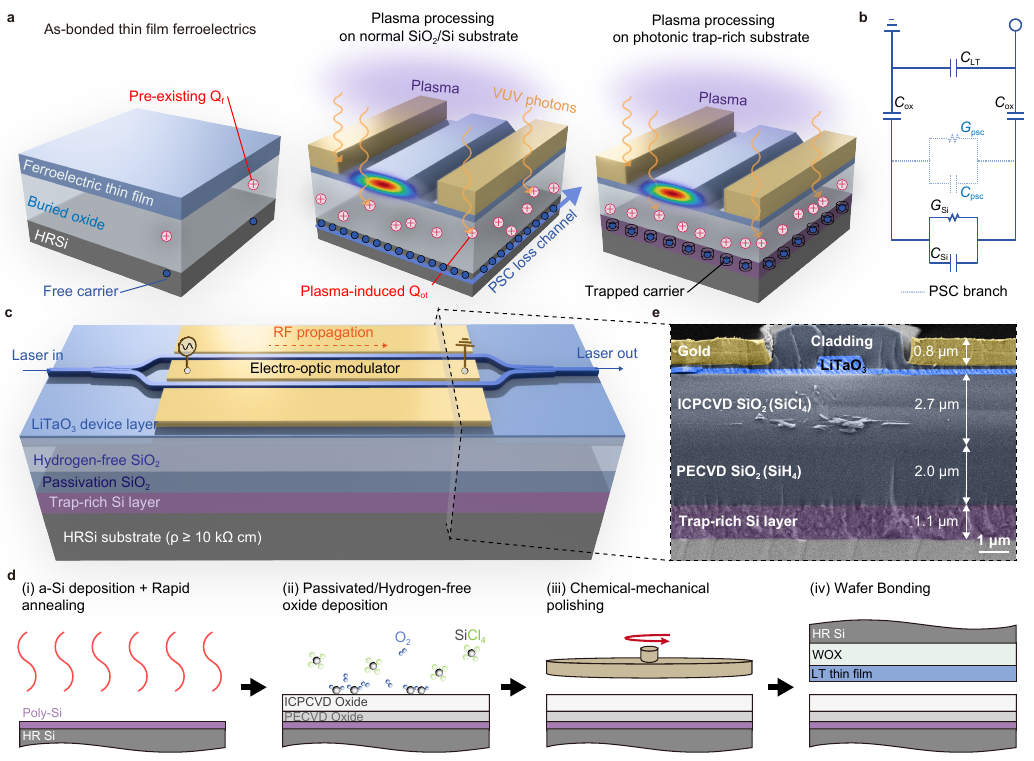}
\caption{\textbf{Trap-rich-engineered LTOI platform enabling low-loss electro-optics.} \textbf{a,} Evolution of interfacial charges during PIC fabrication, showing pre-existing fixed oxide charges \(Q_{\mathrm{f}}\), plasma-induced oxide-trapped charges \(Q_{\mathrm{ot}}\), PSC formation in conventional \ce{SiO2}/Si and carrier trapping in the engineered substrate. \textbf{b,} Simplified lumped-element substrate circuit for a ground--signal--ground electrode. \textbf{c,} Trap-rich LTOI MZM with traveling-wave CPW electrodes. \textbf{d,} Wafer-level fabrication workflow for lithium tantalate on trap-rich substrate, including trap-rich Si formation, multilayer oxide deposition, CMP and hydrophilic wafer bonding. \textbf{e,} False-colored cross-sectional scanning electron micrograph of the corresponding layer stack.}
\label{fig:platform}
\end{figure*}

\subsection{Photonic trap-rich substrate engineering}

	In \LT~EO circuits, the microwave mode supported by the coplanar electrodes extends through the buried oxide and into the silicon substrate, thereby making the substrate stack a critical determinant of RF attenuation.
    Figure~\ref{fig:platform}a illustrates the formation and suppression of the substrate-relevant PSC loss during ferroelectric PIC fabrication.
    Initially, the as-bonded ferroelectric-on-insulator stack contains a low density of pre-existing fixed oxide charges, \(Q_{\mathrm{f}}\).
    VUV photons emitted during subsequent plasma processing can generate electron-hole pairs when their energies exceed the SiO$_2$ bandgap (Supplementary Note 3 and Supplementary Fig.~\ref*{SI-fig:plasma-vuv}).
    Although some pairs recombine, carrier separation allows electrons to be transported out of the oxide, leaving the remaining holes localized and producing additional positive oxide-trapped charges~\cite{yunogami1989radiation}, \(Q_{\mathrm{ot}}\).
    In the conventional wet thermal oxide and HRSi stack beneath the \LT~thin film, these positive charges induce mobile-carrier accumulation at the \(\mathrm{SiO_2/Si}\) interface, forming a conductive PSC sheet that supports interfacial RF currents under microwave excitation.
    The corresponding simplified circuit model in Fig.~\ref{fig:platform}b represents this sheet as an additional lossy shunt branch, \(Y_{\mathrm{PSC}}=G_{\mathrm{PSC}}+i\omega C_{\mathrm{PSC}}\), whose conductance contributes directly to distributed microwave dissipation.
    In the engineered trap-rich platform, the plasma-induced oxide charges may remain, but the defect states in the trap-rich silicon layer immobilize the interfacial carriers before they form the lossy PSC channel.

    From a photonic perspective, the optical mode is confined within the patterned \LT~waveguide core and its surrounding oxide, placing stringent requirements on the optical quality of the buried oxide.
   We therefore developed an adaptable trap-rich silicon substrate for LTOI electro-optics (Fig.~\ref{fig:platform}c) to meet two coupled demands: suppressing substrate-mediated microwave dissipation and preserving an optical-grade buried oxide.
    The resulting wafer stack consists of a \LT~functional layer, a multilayer buried-oxide stack, a trap-rich silicon layer and a HRSi handle substrate.
    Figure~\ref{fig:platform}d outlines the wafer-level process for the substrate-engineered LTOI platform.
    To realize the aforementioned stack, an a-Si layer was deposited on HRSi and thermally stabilized to form the trap-rich Si layer, after which a PECVD passivation oxide using \ce{SiH4} precursor was introduced to protect the electrically active trap states during subsequent inductively coupled plasma chemical vapor deposition (ICPCVD) of a dense, hydrogen-free \ce{SiO2} layer using \ce{SiCl4} and \ce{O2} precursors~\cite{qiu2023hydrogenfree}.
    Chemical-mechanical polishing (CMP) then prepared a bond-ready oxide surface with subnanometer roughness. A 600-nm-thick \LT~film was transferred by hydrophilic bonding followed by thermal annealing~\cite{cai2026heterogeneously}.
    After donor-wafer removal, PICs were fabricated using an established subtractive process~\cite{chen2023highdensity}.
    Fabrication details are provided in Methods and Supplementary Note 2, with the process flow summarized in Supplementary Fig.~\ref*{SI-fig:process-flow}.

    The cross-sectional scanning electron microscope (SEM) image (Fig.~\ref{fig:platform}e) captures one arm of the fabricated MZM, where the 1.5-\um-wide \LT~optical core is positioned between the dense cladding and buried oxide, and adjacent to the 800-nm-thick gold electrodes.
    The resulting structure confirms that our trap-rich substrate can be integrated with the optical-grade oxide stack and the traveling-wave electrode geometry required for low-loss microwave photonics.
	
	\begin{figure*}[t!]
\centering
\includegraphics[width=0.95\textwidth]{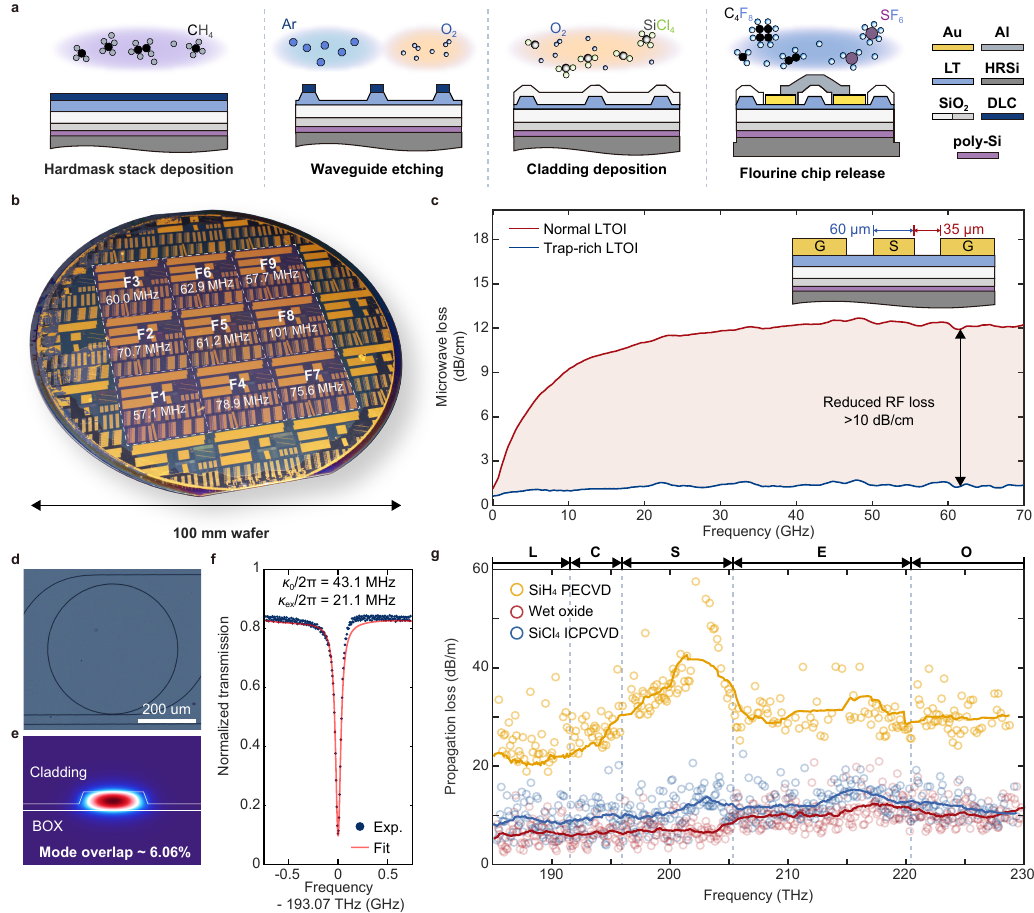}
\caption{\textbf{Simultaneous low microwave and optical loss.} \textbf{a,} Plasma-containing fabrication modules, including \(\mathrm{CH_4}\)-based hard-mask deposition, Ar-ion waveguide etching and \(\mathrm{O_2}\) plasma stripping, \(\mathrm{SiCl_4/O_2}\)-based cladding deposition and fluorine-based chip-release etching. \textbf{b,} Fabricated 100-mm trap-rich LTOI wafer with nine deep-ultraviolet stepper fields. The annotated values indicate average optical intrinsic linewidths extracted from 100 GHz microring resonators in selected fields. Wafer Design ID: D245. \textbf{c,} CPW attenuation on trap-rich and normal-LTOI substrates after the same process. Inset, CPW geometry with a 60~\um{} signal width and a 35~\um{} ground-signal gap on 200 nm \LT~film. \textbf{d,} Optical micrograph image of a 440-\um{}-diameter microring resonator and \textbf{e} the corresponding simulated optical mode profile. \textbf{f,} Representative resonance on the trap-rich LTOI wafer. \textbf{g,} Broadband optical propagation losses derived from microring linewidths for the indicated oxide stacks, with solid curves showing smoothed loss traces.}
\label{fig:loss}
\end{figure*}

\subsection{Wafer-scale trap-rich LTOI photonics combining low microwave and optical losses}

	Scaling thin-film ferroelectric PICs towards volume manufacturing relies on reproducible wafer-level processing~\cite{wang2024lithium}, yet this approach involves repeated plasma-heavy deposition and etching steps.
    As summarized in Fig.~\ref{fig:loss}a, the wafer-scale fabrication flow includes CH$_4$-plasma for diamond-like carbon (DLC) hard-mask deposition, Ar-ion-beam etching and O$_2$-plasma stripping for \LT~waveguide definition, SiCl$_4$/O$_2$ ICPCVD for the hydrogen-free SiO$_2$ cladding as well as fluorine-based oxide etching followed by Bosch deep silicon etching for final chip release.
    Such plasma exposure is known to generate additional \(Q_{\mathrm{ot}}\)~in the buried oxide~\cite{kassabov1988argon,PASKALEVA20002033,awadelkarim1996plasma}, with contributions potentially accumulating across successive PIC fabrication steps.
    To isolate substrate-dependent effects, both trap-rich and normal-LTOI (600 nm~\LT/ 4.7~\um{} wet thermal oxide/ 700~\um{} HRSi) wafers were subjected to the same fabrication flow (Methods and Supplementary Fig.~\ref*{SI-fig:process-flow}).
    Figure~\ref{fig:loss}b shows a photograph of a finalized 100-mm trap-rich LTOI wafer, with the transferred~\LT~ film hosting EO and passive photonic test structures across all nine DUV stepper fields.
   As demonstrated in Supplementary Note 4 and Supplementary Fig.~\ref*{SI-fig:edge-trimming}, a dedicated edge trimming step can further increase the effective thin-film transfer area.

    We first compared the room-temperature microwave losses of 5-mm-long reference CPW transmission lines on the trap-rich and normal-LTOI substrates using calibrated \(S_{21}\) measurements (Fig.~\ref{fig:loss}c; see Methods).
    For the CPWs located in the center of the wafer (Field 5), the loss of the trap-rich device remained below 1.6~dB/cm up to 70 GHz, whereas that of the normal-LTOI reference increased rapidly and leveled off near 12~dB/cm at high frequencies.
   Specifically, the engineered trap-rich substrate stack reduced the microwave attenuation by 10.8~dB/cm at 60 GHz relative to the normal-LTOI reference.
    This broadband reduction indicates the trap-rich substrate as the primary contributor to the improved microwave behavior, consistent with suppression of the PSC branch in our aforementioned circuit model.

    To determine whether this substrate engineering compromises optical performance, we characterized microring resonators fabricated on the same trap-rich LTOI platform.
    Figure~\ref{fig:loss}d shows a microring resonator with a diameter of 440~\um and a free-spectral-range (FSR) of 100 GHz.
    The corresponding simulated cross-sectional mode profile in Fig.~\ref{fig:loss}e displays a mode overlap of \(6.06\%\) with the buried oxide.
    Although most of the optical mode remains confined to the \LT~waveguide core, this non-negligible overlap makes the measured resonator linewidth sensitive to the optical absorption in the oxide part.
    A representative resonance near 193.07 THz, measured in the central wafer field (Field 5), is shown in Fig.~\ref{fig:loss}f. The fit to the resonance yields an intrinsic linewidth of \(\kappa_0/2\pi = 43.1~\mathrm{MHz}\) and an external coupling rate of \(\kappa_{\mathrm{ex}}/2\pi = 21.1~\mathrm{MHz}\), corresponding to an intrinsic quality factor of approximately \(4.5 \times 10^6\) and propagation loss of around 9 dB/m.
    The broadband optical response was further assessed by converting the fitted intrinsic linewidths into propagation losses for resonances spanning the L, C, S, E and O telecommunication bands, as summarized in Fig.~\ref{fig:loss}g.
    Microring resonators incorporating the SiCl$_4$-ICPCVD oxide stack exhibit propagation losses of approximately 10~dB/m across this wide spectral window, comparable to those of the conventional wet-oxide reference.
    By comparison, reference resonators with SiH$_4$-PECVD buried oxide show substantially higher propagation losses, together with a pronounced loss peak in the shorter-wavelength bands. These features are consistent with hydrogen- or hydroxyl-related absorption in silica~\cite{qiu2023hydrogenfree,wu2023hydroxyl}.
    We assessed wafer-level optical-loss uniformity across five representative DUV fields and measured field-averaged intrinsic linewidths ranging from 57.1 to 75.6 MHz (Fig.~\ref{fig:loss}b; Supplementary Note 10 and Supplementary Fig.~\ref*{SI-fig:wafer-analysis}).
	
	\begin{figure*}[t!]
\centering
\includegraphics[width=0.93\textwidth]{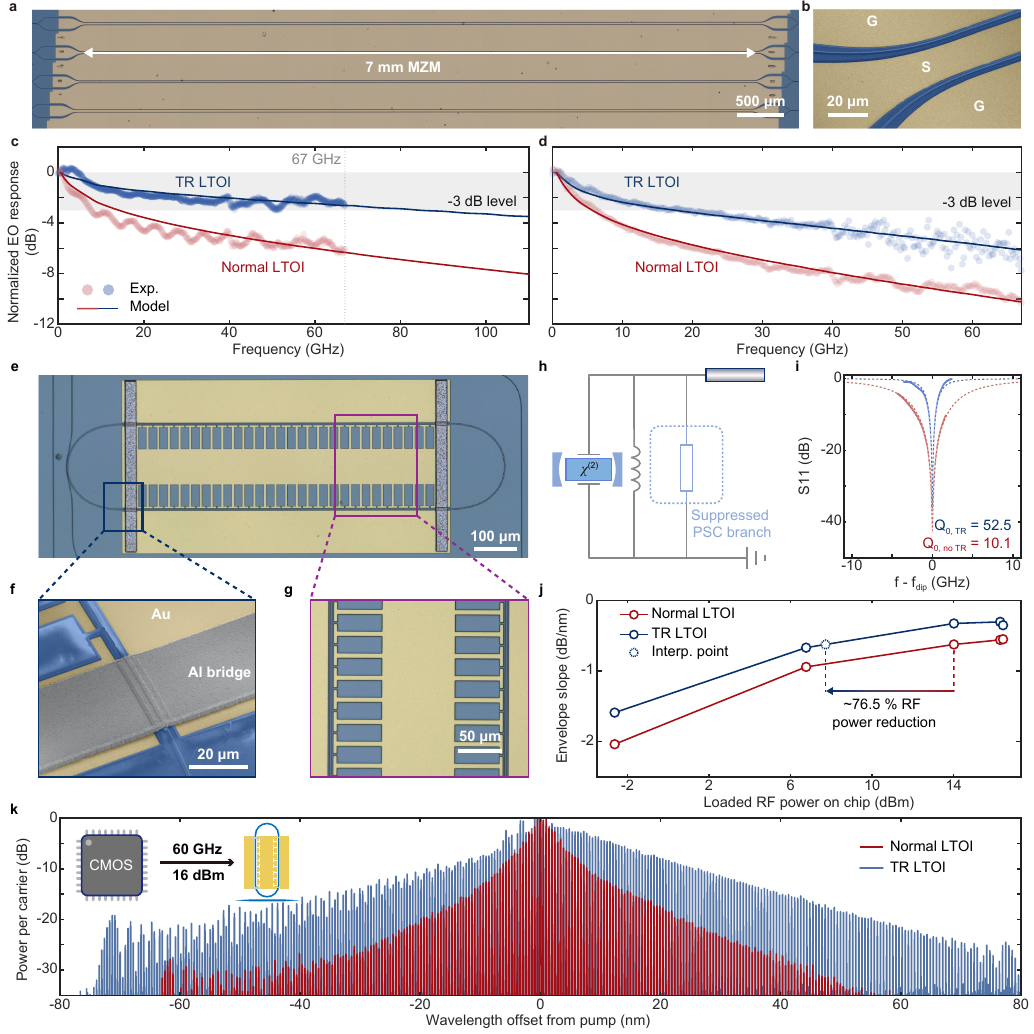}
\caption{\textbf{Electro-optic device performance.} \textbf{a,} Optical micrograph of a 7-mm-long MZM. \textbf{b,} False-colored SEM image of the ground-signal-ground electrode region. \textbf{c} and \textbf{d,} Measured and modeled EO responses of the 7-mm-long MZMs with 5~\um{} ground-signal electrode gap and 17-mm-long MZMs with 6~\um{} ground-signal electrode gap, respectively, fabricated on trap-rich (TR) and normal LTOI substrates under the same process. \textbf{e,} Optical micrograph of the triply resonant EO comb devices. \textbf{f} and \textbf{g,} Enlarged Al bridge and capacitive-electrode regions. \textbf{h,} Simplified schematic of the triply resonant EO comb architecture. \textbf{i,} Microwave reflection spectra and fitted intrinsic quality factors. \textbf{j,} Fitted comb-envelope slope versus calibrated on-chip RF power. \textbf{k,} Comb spectra at the same 16-dBm on-chip RF power.}
\label{fig:EO}
\end{figure*}

\subsection{Modulator and EO comb performance}

	We compared the EO performance of 7-mm- and 17-mm-long traveling-wave MZMs fabricated on trap-rich and normal-LTOI substrates (Fig.~\ref{fig:EO}a and b).
    Within each length pair, the trap-rich and normal-LTOI devices shared an identical layout, with the 7-mm-long MZMs using a 5~\um{} ground-signal gap and the 17-mm-long MZMs using a 6~\um{} gap (Supplementary Note 5 and Supplementary Fig.~\ref*{SI-fig:mzm-simulation}).
     Figures~\ref{fig:EO}c and d show the corresponding measured and modeled EO responses.
    For the 7-mm-long devices, the EO response of the trap-rich device was consistently higher than that of the normal-LTOI reference across the measured frequency range, with the latter crossing the -3 dB level near 10 GHz.
    The periodic ripple observed in Fig.~\ref{fig:EO}c reveals the coherent reflections associated with non-adiabatic pad-to-CPW microwave transitions.
    As detailed in Supplementary Note 7 and Supplementary Fig.~\ref*{SI-fig:si-eo-model-analysis}, we replaced the oscillatory effective impedance with its smooth background to assess the broadband EO capability under an idealized microwave interconnection.
    The resulting projection yielded a 3-dB EO bandwidth of approximately 85 GHz.
    For the 17-mm-long MZMs, the trap-rich and normal-LTOI devices exhibited \(3\)-dB EO bandwidths of approximately 16.2 and 6.3 GHz, respectively.
    We measured comparable normalized EO responses for identical MZMs in five representative DUV fields (Supplementary Fig.~\ref*{SI-fig:wafer-analysis}), further demonstrating reproducible device performance across the fabricated wafer.
    The measured half-wave voltages were 4.55 V for the 7-mm-long MZM with a 5~\um{} electrode gap and 2.47 V for the 17-mm-long MZM with a 6~\um{} electrode gap (Supplementary Note 6 and Supplementary Fig.~\ref*{SI-fig:vpi}).

    Next, we compared resonant electro-optic frequency-comb generation in devices fabricated on trap-rich and normal-LTOI substrates.
    In Fig.~\ref{fig:EO}e, each device co-integrates a 61-GHz-FSR optical racetrack microresonator with a capacitively loaded, $\lambda/2$ coplanar-waveguide microwave resonator.
   The enlarged views highlight the Al-bridge short terminations at both ends that form the $\lambda/2$ resonator (Fig.~\ref{fig:EO}f), as well as the capacitive electrode section used to tune the microwave phase matching and reduce microwave loss (Fig.~\ref{fig:EO}g).
    The microwave resonator was designed to operate near 60~GHz, matching the optical FSR. The resulting triply resonant architecture (Fig.~\ref{fig:EO}h), in which the trap-rich substrate suppresses the PSC-related shunt loss, resonantly enhances both the optical sidebands and the intracavity microwave field, thereby strengthening the EO coupling and enabling broader EO-comb generation.
    Figure~\ref{fig:EO}i shows the measured microwave reflection spectra, \(S_{11}\), and the corresponding fits. The fitted intrinsic microwave quality factor was \(Q_0=52.5\) at 61.08 GHz for the trap-rich device, approximately \(5.2\) times that of the normal-LTOI reference (\(Q_0=10.1\) at 61.83 GHz, reflecting substantially reduced intrinsic microwave dissipation in the trap-rich resonator. The fitting procedure is detailed in Supplementary Note 8 and Supplementary Fig.~\ref*{SI-fig:mw-resonance}.
    Figure~\ref{fig:EO}j shows the comb-envelope slopes as a function of calibrated on-chip RF input power (see Supplementary Note 9 and Supplementary Fig.~\ref*{SI-fig:comb-envelope} for fitting details), demonstrating that the trap-rich device produces a flatter comb envelope at a given input power.
   At the matched comb-envelope shape, represented here by a fitted slope of \(-0.625~\mathrm{dB\,nm^{-1}}\), the normal-LTOI device required 14.01 dBm, whereas linear interpolation of the trap-rich data yielded a required RF power of 7.72 dBm, corresponding to a 6.29 dB power reduction.
    Because both devices employ the same triply resonant architecture, this near 76.5\% power reduction is consistent with the approximately 5.2-fold enhancement in \(Q_0\) of the trap-rich microwave resonator.
   The improvement is also evident in the representative spectra measured at the maximum calibrated on-chip RF power of 16~dBm, where the trap-rich device produces a broader comb spectrum than the normal-LTOI device (Fig.~\ref{fig:EO}k).
    The substrate-level improvement therefore provides an efficiency gain beyond the enhancement already afforded by the triply resonant structure relative to devices using a non-resonant electrode~\cite{zhang2025ultrabroadband}.

	\begin{figure*}[t!]
\centering
\includegraphics[width=0.95\textwidth]{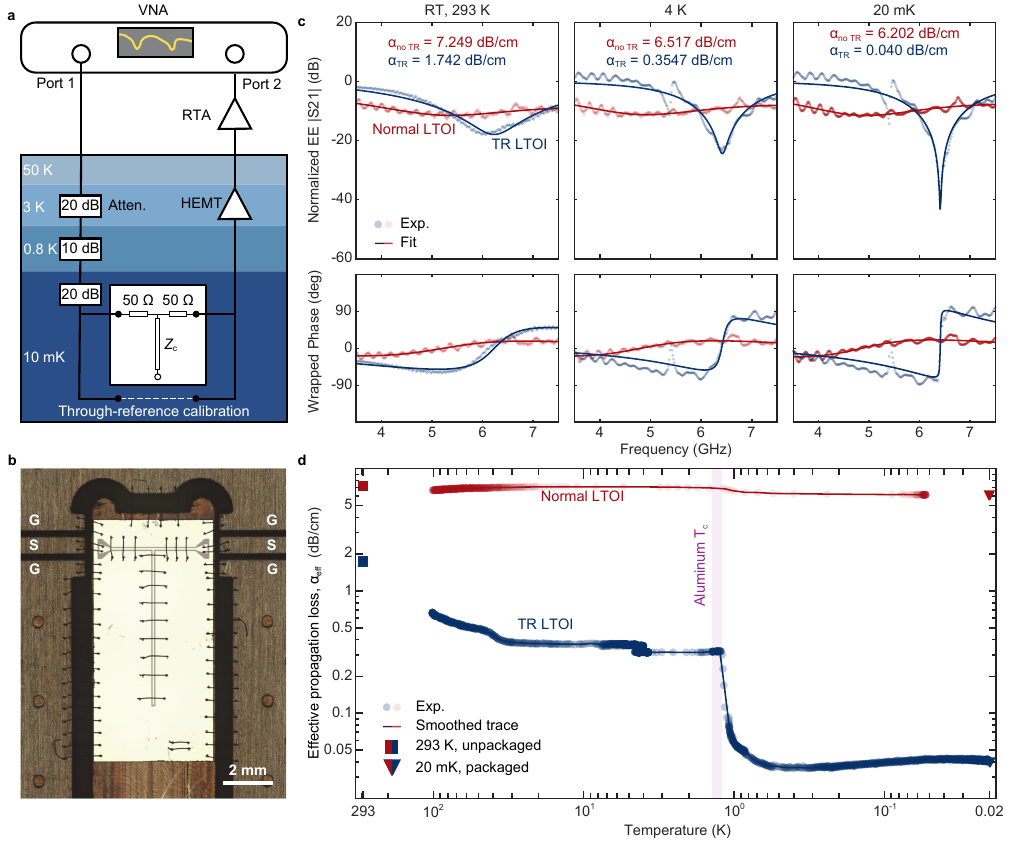}
\caption{\textbf{Cryogenic microwave-loss comparison.} \textbf{a,} Schematic of the two-port cryogenic measurement setup. HEMT, High-electron-mobility-transistor; RTA, Room-temperature amplifier. \textbf{b,} Photograph of the packaged microwave test chip. \textbf{c,} Representative normalized \(S_{21}\) magnitudes (upper panels) and wrapped phases (lower panels) measured at 293~K, 4~K and 20~mK. \textbf{d,} Effective propagation loss, \(\alpha_{\mathrm{eff}}\), as a function of temperature. Squares and triangles denote the values at 293~K and 20~mK, respectively. The shaded region marks the expected superconducting transition of the Al electrodes near \(T_c\approx1.2~\mathrm{K}\).}
\label{fig:cryo}
\end{figure*}

\subsection{Cryogenic device performance}

        Cryogenic cooling suppresses thermally activated bulk conduction in HRSi, but this bulk-carrier freeze-out does not eliminate the PSC-induced microwave dissipation.
    	In bulk HRSi, conductivity is governed by competing changes in carrier mobility and density.
    	Reduced phonon scattering can initially increase mobility, whereas dopant freeze-out at lower temperatures depletes mobile carriers and restores high bulk resistivity~\cite{krupka2006measurements}.
    	However, once fixed positive oxide charges are present, their induced silicon surface inversion layers can maintain the band bending and conduction at cryogenic temperatures~\cite{shiue1976studies}.
    	Accordingly, recent measurements have observed prominent substrate loss and a continuing benefit from interface-passivation or trap-rich layers in plasma-processed LNOI and RF-SOI platforms~\cite{berlingard2023rf,shen2024parasitic}.
    	We therefore evaluated whether the microwave-loss advantage of trap-rich LTOI persisted throughout cryogenic cooling, including below the superconducting transition of the electrodes.
    	For this purpose, we designed and fabricated the Al T-resonators~\cite{peterson2001cpw} with a fundamental resonance near 6~GHz to probe microwave loss using cryogenic $S_{21}$ measurements (Fig.~\ref{fig:cryo}a and b).
        After through-reference calibration and normalization, the complex \(S_{21}\) spectrum measured at each temperature was fitted using the distributed shunt-stub model, yielding the corresponding effective propagation loss \(\alpha_{\mathrm{eff}}(T)\) (Methods; Supplementary Note 11 and Supplementary Fig.~\ref*{SI-fig:t-resonator}).
    	Representative \(S_{21}\) spectra and fits at 293~K, 4~K and 20~mK are shown in Fig.~\ref{fig:cryo}c, while Fig.~\ref{fig:cryo}d reveals a pronounced difference in effective propagation loss between the trap-rich device and the normal-LTOI reference throughout the measured temperature range. At 293~K, \(\alpha_{\mathrm{eff}}\) is 1.742~dB/cm for trap-rich LTOI and 7.249~dB/cm for normal LTOI; at the base temperature, the corresponding values are 0.040 and 6.202~dB/cm, yielding a loss contrast of 6.162~dB/cm.
   		With both devices employing the same Al electrodes, the change in \(\alpha_{\mathrm{eff}}\) near \(T_c\approx1.2~\mathrm{K}\) coincides with the expected superconducting transition.
   		However, the normal-LTOI reference without the trap-rich layer continued to exhibit excessive microwave loss, largely offsetting the benefit of the superconducting metallization.
		
	\subsection{Discussion}
		To summarize, we developed a 100 mm trap-rich LTOI substrate that suppresses parasitic surface conduction (PSC) in room-temperature electro-optic photonic integrated circuits while preserving low optical loss. The substrate combines a carrier-trapping silicon layer with a hydrogen-free SiCl$_4$-based ICPCVD oxide, providing a dense, bonding-ready and photonics-compatible trap-rich stack.
        To the best of our knowledge, this work is the first wafer-scale demonstration of a trap-rich substrate engineered for thin-film ferroelectric photonics.
        Across passive and active devices, the engineered substrate enabled broadband traveling-wave modulation and more power-efficient resonant electro-optic comb generation.
        The microwave-loss advantage of the engineered substrate persisted down to 20~mK, including below the superconducting transition of the Al metallization.
        Given the demonstrated device performance, our photonic trap-rich LTOI platform offers a process-robust approach to scalable ferroelectric electro-optics within the silicon manufacturing ecosystem, even as future fabrication flows continue to grow in complexity.

        More broadly, by tailoring the bonding chemistry and thermal budget to each material, this wafer-level substrate-engineering strategy can be adapted to other thin-film ferroelectric platforms, including \LN, barium titanate (BTO) and lead zirconate titanate (PZT).
        By combining low optical and microwave loss, compatibility with wafer-scale processing and cryogenic robustness, this platform provides a foundation for high-density optical interconnects~\cite{han2025exploring}, cryogenic microwave-to-optical transducers~\cite{han2021microwave,xu2021bidirectional} and energy-efficient ultrabroadband electro-optic comb generators~\cite{zhang2025ultrabroadband}.

	\section*{Methods}
	
	\noindent\textbf{Trap-rich substrate formation and wafer bonding:} A 1.1-\um{} a-Si layer was deposited on the HRSi handle wafer by low-pressure chemical vapor deposition and annealed at 900$^\circ$C for 2~min to form a thermally stable carrier-trapping poly-Si layer. A 2-\um{} passivation oxide was deposited using low-plasma-power plasma-enhanced chemical vapor deposition (PECVD), followed by 3.2~\um{} of hydrogen-free oxide deposited by inductively coupled plasma chemical vapor deposition with SiCl$_4$ and O$_2$ precursors.
    The first PECVD oxide was introduced to limit direct exposure of the carrier-trapping interface to the aggressive chlorine-rich plasma. Chemical-mechanical polishing set the total buried-oxide thickness to 4.7~\um{} and produced a bond-ready surface with a root-mean-square roughness of 0.208 nm (Supplementary Note 1).
    A 600-nm \LT\ film was transferred to the engineered handle by O$_2$-plasma-activated hydrophilic bonding and annealing at 450$^\circ$C for more than 5~hours.
    During this bonding step, initial contact between the two wafers was established through van der Waals interactions, and was further strengthened by subsequent annealing to convert silanol groups into covalent Si--O--Si bonds.
    The donor silicon was thinned by mechanical grinding and then completely removed by deep reactive-ion etching.
    Fluorine-based dry etching and buffered hydrofluoric acid removed the donor-side oxide to expose the transferred \LT\ film.
	
	\noindent\textbf{Photonic device fabrication:} Photonic integrated circuits were fabricated on the bonded trap-rich LTOI wafer using a DLC hard-mask process~\cite{chen2023highdensity}.
   Waveguide patterns were defined by deep ultraviolet (DUV) lithography (ASML PAS 5500 stepper, using a 248 nm KrF laser source) and transferred into the \(600\)-nm-thick \LT~film by argon ion-beam etching, followed by alkaline wet cleaning to remove sidewall redeposition and \ch{O2}-plasma stripping of the residual hard mask. A 1.5-\um-thick hydrogen-free \ce{SiO2} cladding was subsequently deposited using the same \ce{SiCl4}-ICPCVD process as for the buried oxide. Electrode patterns were defined by a second DUV lithography step and transferred through the cladding by fluorine-based dry etching. Following an approximately 300 nm buffered-hydrofluoric-acid undercut, 20 nm Al/800 nm Au electrodes were formed by electron-beam evaporation and lift-off. For the microwave resonators, 1.8-\um-thick Al air bridges were added by maskless UV lithography, electron-beam evaporation and selective wet etching.
   The chips were defined with a final maskless UV lithography step followed by first an oxide etch and then a \(240~\um\) deep Si etch.
   Finally, the chips were released by wafer-backside grinding.
	
	\noindent\textbf{Optical loss characterization:} The optical propagation loss was characterized by optical-frequency-comb-calibrated laser spectroscopy of ring resonators and estimated from the fitted intrinsic cavity linewidths.
    Three tunable lasers (SANTEC TSL-510, TSL-550 and TSL-570) were used to continuously sweep the wavelength range from 1260 nm to 1630 nm, with the absolute wavelength of each sweep calibrated using a gas cell and a locked optical frequency comb (MENLO Systems FC-1500).
    Both the amplitude and phase of the cavity transmission were measured; the phase information was obtained using a Mach--Zehnder interferometer and extracted via a Hilbert transform.
    The complex transmission was fitted to extract separately the intrinsic cavity linewidth, $\kappa_0$, and external coupling rate, $\kappa_{\mathrm{ex}}$, from which the optical propagation loss was determined. Devices fabricated on the trap-rich and conventional LTOI wafers used the same optical mask (stepper reticle, internal ID D245), ensuring an identical optical design.

    \noindent\textbf{Microwave and EO measurements:} Microwave attenuation was measured on 5-mm CPW transmission lines with a vector network analyzer (Rohde \& Schwarz ZNA67).
    A two-port through--open--short--match calibration on an MW-CS5 calibration substrate placed the reference planes at the tips of the two electrical probes. Propagation loss was extracted from the calibrated $S_{21}$ spectra of geometry-matched lines on trap-rich and normal-LTOI substrates.

    For $S_{\mathrm{EO}}$ measurements, a TOSM calibration with a commercial calibration module placed the electrical input reference plane at the coaxial interface immediately before the probe.
    Because the EO-response measurement involved only one electrical probe at the input port, the probe loss was de-embedded using supplier-provided loss data.
    The modulated optical signal was detected using a fast \SI{110}{\giga\hertz} photodiode (Finisar XPDV4120R), whose optical-to-electrical frequency response was independently characterized using two-laser beat-note measurements and subsequently de-embedded from the measured EO response.

    \noindent\textbf{Cryogenic measurements:} Geometry-matched CPW T resonators were fabricated from 500-nm Al on the two LTOI substrates. Measurements were performed in a Bluefors SD dilution refrigerator. The vector-network-analyzer drive was attenuated at successive cryostat stages; the transmitted signal was amplified by a cryogenic high-electron-mobility-transistor amplifier and a room-temperature amplifier. A separately measured device-free through path was used for complex de-embedding.

	\section*{Data and code availability}
	The data and processing code used in this research will be deposited in the public repository Zenodo and made publicly available upon publication of the manuscript.

\section*{Acknowledgements}
	This work was supported by the National Key R\&D Program of the Ministry of Science and Technology of China (grant No. 2022YFA1404601) and the National Natural Science Foundation of China (grant No. 62293521), awarded to X.O; the European Research Council (ERC) under the Horizon Europe research and innovation programme, grant agreement No. 101167540 (ATHENS), awarded to T.J.K.; ; and the Independent Research Fund Denmark (grant No. 5249-00033B), awarded to M.A. We thank Yang Chen and Xinjian Ke of Shanghai Novel Si Integration Technology (NSIT) for technical guidance on wafer bonding and for providing the commercial LTOI wafers used in this work.
    We thank Nikolai Kuznetsov for assistance with photography.
    The LTOI wafers were fabricated at NSIT and the Shanghai Institute of Microsystem and Information Technology, Chinese Academy of Sciences.
    The photonic samples were fabricated in the EPFL Center of MicroNano Technology.
	\section*{Author contributions}
	J.C., M.A., and C.W. developed the photonic trap-rich silicon substrate. J.C., M.A., X.J., Y.Z., and Z.Q. fabricated the trap-rich LTOI photonic integrated circuits. J.C., J.Z., M.L., and S.Z. designed and characterized the EO devices and cryogenic microwave resonators. J.C., J.Z., M.A., and S.Z. analyzed the data. T.J.K. and X.O. supervised all aspects of the project. J.C., J.Z., and M.A. wrote the manuscript with input from all authors. All authors discussed the results and reviewed the manuscript.
	
	\section*{Competing interests}
	T.J.K. is a co-founder and shareholder of Luxtelligence SA, a foundry commercializing \LN~and \LT~photonics, as well as DEEPLIGHT SA, a start-up commercializing PIC-based frequency-agile low-noise lasers.
    The other authors declare no competing interests.
	
	\section*{Materials \& Correspondence}
	Correspondence and requests for materials should be addressed to X.O and T.J.K.
	
	\hypersetup{urlcolor=black}
\bibliographystyle{apsrev4-2}
\bibliography{references}
\end{document}


\title{Supplementary Information for: Unlocking scalable manufacturing of high-performance integrated electro-optic circuits via substrate engineering}
	
	\author{Jiachen~Cai$^{1,2\dagger}$,
	Junyin~Zhang$^{2\dagger}$,
	Marcus~Albrechtsen$^{2\dagger}$,
	Shuhang~Zheng$^{2}$,
	Chengli~Wang$^{1}$,
	Mengxin~Lin$^{2}$,
	Xinru~Ji$^{2}$,
	Yichi~Zhang$^{2}$,
	Zheru~Qiu$^{2}$,
	Xin~Ou$^{1\ast}$,
	and Tobias J. Kippenberg$^{2,3\ast}$}
	\affiliation{
		 State Key Laboratory of Materials for Integrated Circuits, Shanghai Institute of Microsystem and Information Technology, Chinese Academy of Sciences, Shanghai, China\\
		 Institute of Physics, Swiss Federal Institute of Technology Lausanne (EPFL), CH-1015 Lausanne, Switzerland\\
		 Institute of Electrical and Micro Engineering, Swiss Federal Institute of Technology Lausanne (EPFL), CH-1015 Lausanne, Switzerland\\
		\normalsize{$^\ast$Corresponding authors. Email: ouxin@mail.sim.ac.cn; tobias.kippenberg@epfl.ch}\\
        \normalsize{$^\dagger$These authors contributed equally to this work.\\}}
	
\setcounter{equation}{0}
\setcounter{figure}{0}
\setcounter{table}{0}

\setcounter{subsection}{0}
\setcounter{section}{0}
\setcounter{secnumdepth}{3}
	
	\maketitle
{
\renewcommand{\tocname}{Contents for Supplementary Notes}
\hypersetup{linkcolor=blue}
\tableofcontents
}
\newpage

\renewcommand{\thesection}{SN\arabic{section}}
\section{Pre-bonding roughness quantification}

\begin{figure}[!htbp]
  \centering
  \includegraphics[width=0.78\linewidth]{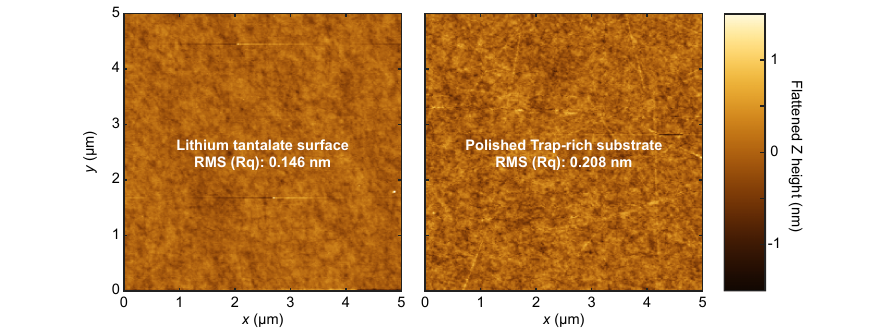}
  \caption{\textbf{Atomic force microscopy of the wafer surfaces to be bonded.}
  Plane-flattened AFM height maps measured over \(5~\um{}\times~5~\um{}\) areas of the lithium tantalate surface and the CMP-polished oxide surface of the trap-rich handle wafer, corresponding to RMS roughness values of \(R_q=0.146\) nm and \(0.208\) nm, respectively.}
  \label{fig:surface-rough}
\end{figure}

Before wafer bonding, the surface of the lithium-tantalate-on-insulator (LTOI) wafer and the oxide surface of the trap-rich substrate were characterized using an atomic force microscope (Park Systems FX200) in non-contact mode over \(5~\um \times 5~\um\) areas, where the oxide surface was measured after chemical-mechanical polishing (CMP). The measured root-mean-square (RMS) roughness values are $R\text{q}=\SI{0.146}{\nano\meter}$ and \SI{0.208}{\nano\meter}, respectively (\fref{fig:surface-rough}). Importantly, both surfaces therefore exhibited sub-nanometre roughness, satisfying the hydrophilic wafer bonding roughness requirement, $R\text{q}\leq0.5$~nm~\cite{cai2026heterogeneously}.

\section{Fabrication process flow}

\begin{figure}[!htbp]
  \centering
  \includegraphics[width=0.96\linewidth]{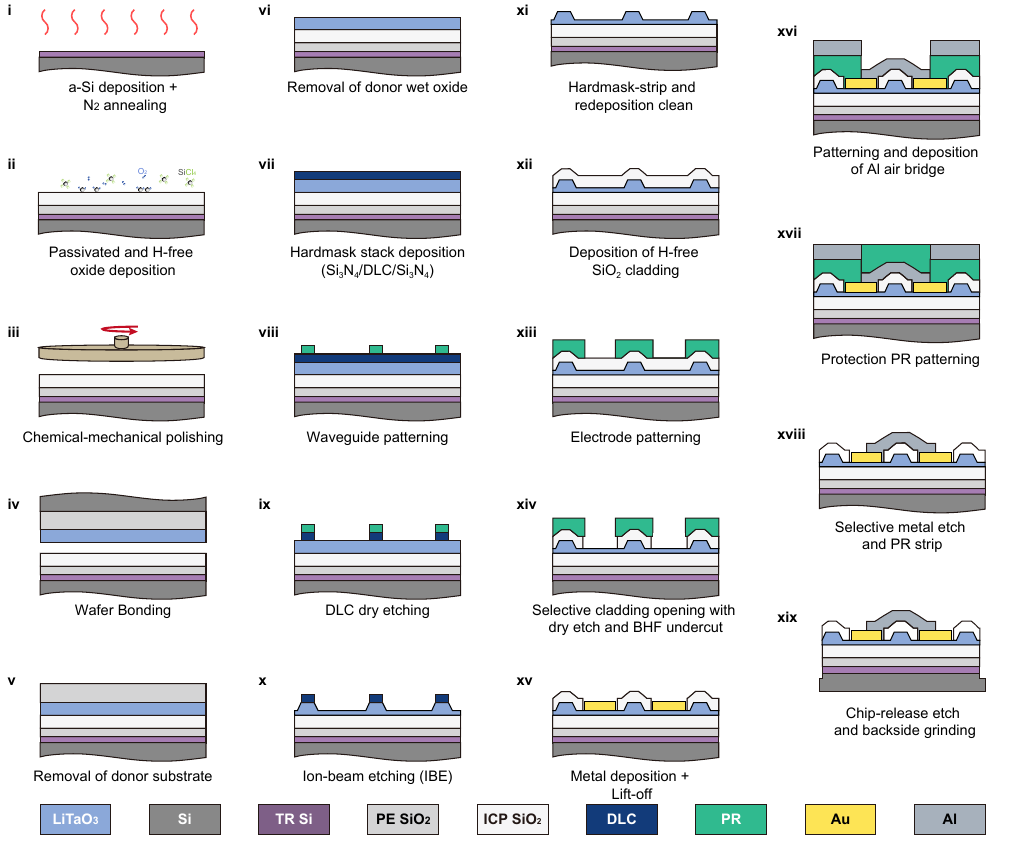}
  \caption{\textbf{Full process flow for trap-rich LTOI photonic integrated circuits.}
  The substrate preparation (steps (i)-(vi)) process comprises formation of the 1.1~\um~a-Si trap-rich layer and bilayer \ce{SiO2} stack, CMP, wafer bonding, and subsequent removal of the donor silicon handle and oxide. The device fabrication includes DLC-assisted \LT\ waveguide patterning (steps (vii)-(xi)), hydrogen-free cladding deposition (xii), electrode and metal-bridge formation (steps (xiii)-(xviii)), and finally chip release (xix).}
  \label{fig:process-flow}
\end{figure}

The complete substrate-preparation and PIC-fabrication sequence is summarized in \fref{fig:process-flow}.
A 1.1-\um{} amorphous-silicon layer was first deposited on the HRSi handle wafer by low-pressure chemical vapor deposition and annealed at 900$^\circ\mathrm{C}$ for 2~min to form a thermally stable carrier-trapping poly-Si layer. A 2-\um{} passivation oxide was deposited using low-plasma-power plasma-enhanced chemical vapor deposition (PECVD), followed by 3.2~\um{} of hydrogen-free oxide deposited by inductively coupled plasma chemical vapor deposition with SiCl$_4$ and O$_2$ precursors.
The first PECVD oxide was introduced to limit direct exposure of the carrier-trapping interface to the aggressive chlorine-rich plasma. Chemical-mechanical polishing set the total buried-oxide thickness to 4.7~\um{} and produced a bond-ready surface.
A 600-nm-thick \LT\ film was transferred to the engineered handle by O$_2$-plasma-activated hydrophilic bonding and annealing at 450$^\circ\mathrm{C}$ for more than 5~hours.
During this bonding step, initial contact between the two wafers was established through van der Waals interactions, and was further strengthened by subsequent annealing to convert silanol groups into covalent molecular bonds.
After that, the donor silicon was thinned by mechanical grinding and then completely removed by deep reactive-ion etching.
Fluorine-based dry etching and buffered hydrofluoric acid (BHF) removed the donor-side wet oxide to expose the transferred \LT\ film.

Following thin-film transfer, ferroelectric photonic integrated circuits were fabricated using a diamond-like carbon (DLC) hard-mask process~\cite{chen2023highdensity}. A 40~nm \ce{Si3N4} film was deposited by PECVD (Plasma-Therm Corial 250L) to promote adhesion followed by a \SI{500}{\nano\meter} DLC hardmask by PECVD using a methane precursor (Santech SI 500). Then a thin 60~nm \SiN~film was deposited on the DLC film as a hardmask for the subsequent oxygen DLC etching, which concludes the hardmask trilayer stack. The waveguide patterns were defined using a deep ultra-violet (DUV) stepper (ASML PAS 5500/350C) and M108Y photoresist with a bottom anti-reflective coating (BARC) and transferred into the hardmask stack using four etching steps in the same tool, a BARC etch, a \ce{Si3N4} etch, a DLC etch and finally a second \ce{Si3N4} etch using fluorine- and oxygen-based plasma etching at low pressure. Then the \LT\ thin film was etched by Ar-based ion-beam etching (IBE), using the high selectivity offered by the DLC hardmask, followed by wet etching in an alkaline solution to remove the redeposited residues sputtered on the waveguide sidewalls.

A short oxygen plasma process was then used to strip the residual DLC hardmask, completing the waveguide fabrication. Next, a 1.5~\um{} H-free \ce{SiO2} cladding was deposited (step (xii)) using the same \ce{SiCl4}-ICPCVD process used for the bottom oxide deposition (step (ii)). The electrode layout was defined by a second DUV lithography step on the top of the cladding and was transferred through the top \ce{SiO2} cladding-layer with fluorine-based ICP dry etching. Then a BHF wet-etching step was carried out to form a $\sim\SI{300}{\nano\meter}$ undercut, and a bilayer metal stack formed of \SI{20}{\nano\meter} aluminum for adhesion and \SI{800}{\nano\meter} gold was deposited with e-beam evaporation and lift-off. To ensure proper operation of the microwave resonator a metal air bridge connection was defined across the optical waveguides. Specifically, 1.8~\um{} aluminum bridges were patterned using a maskless aligner (MLA, Heidelberg Instruments MLA150) and deposited by e-beam evaporation, followed by an additional MLA exposure step to selectively protect the bridge portions before aluminum wet etching. After the removal of the photoresist, the final chip-release processes involved hardmask definition of chip boundaries, dry etching in \SiO~using fluorine-based chemistry, deep silicon etching using the Bosch process and backside wafer grinding.

\section{Plasma VUV emission and oxide charging}

\begin{figure}[!htbp]
\centering
\includegraphics[width=0.98\linewidth]{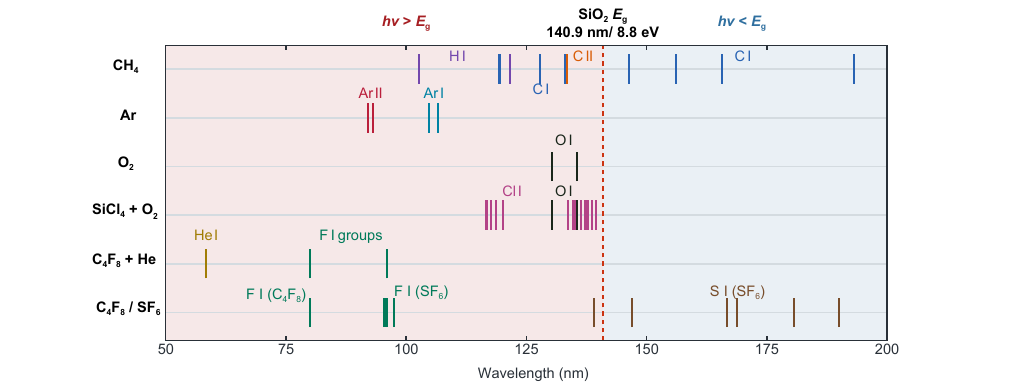}
\caption{\textbf{Representative plasma VUV emission lines relative to the \ce{SiO2} bandgap.}
Vertical markers indicate selected emission wavelengths associated with the indicated plasma chemistries, with a dashed line at 140.9 nm corresponding to a reference \ce{SiO2} bandgap of 8.8 eV.}
\label{fig:plasma-vuv}
\end{figure}

Representative VUV emission wavelengths associated with plasma chemistries are listed in \fref{fig:plasma-vuv}, including \ce{CH4} for DLC hard-mask deposition~\cite{pang1987electron}; Ar for ion-beam etching of lithium-tantalate waveguides~\cite{espinho2017vacuum} and \ce{O2} for DLC hard-mask patterning, removal and surface activation before wafer bonding~\cite{fiebrandt2020oxygen}; \ce{SiCl4}/\ce{O2} for hydrogen-free buried-oxide and  dense cladding deposition by ICPCVD~\cite{woodworth1999absolute}; \ce{C4F8}/He for oxide dry etching~\cite{woodworth2001absolute, samukawa2001onwafer}; \ce{C4F8}/\ce{SF6} for Bosch deep silicon etching during chip release~\cite{li1996vacuum, roque1991production}.

Several chemistries exhibit emission lines below approximately 140 nm, corresponding to photon energies sufficient to generate electron–hole pairs in the oxide.
The comparison illustrates the energetic basis for plasma-induced oxide charging, which does not quantify the photon flux or trapped-charge contribution of individual fabrication steps.

\section{Edge trimming for improving thin-film transfer performance}

\begin{figure}[!htbp]
  \centering
  \includegraphics[width=0.98\linewidth]{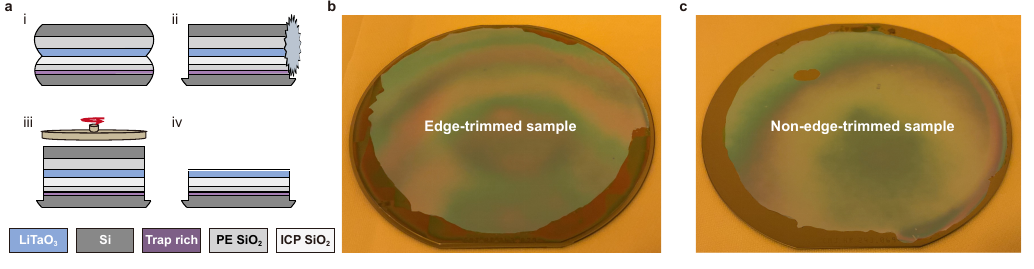}
  \caption{\textbf{Edge trimming for improved thin-film transfer.}
  \textbf{a}, Schematic of the edge-trimming procedure introduced before donor-side substrate-removal to suppress grinding-induced edge damage. \textbf{b,c}, Photographs of the transferred films obtained with (\textbf{b}) and without (\textbf{c}) edge trimming. The underlying film is preserved with edge-trimming, protecting the Si handle wafer during subsequent processing. Regions with poor bonding at the very edge will still delaminate and can be optimized with further developments.}
  \label{fig:edge-trimming}
\end{figure}

Wafer bonding requires a strong and intimate bond across the main wafer area. However, the \LT\ thin film can delaminate near the wafer perimeter during donor-side silicon thinning, particularly when backside grinding concentrates mechanical stress and initiates film peel-off. To mitigate this problem, we utilized an edge-trimming step before the removal of the donor silicon, as illustrated in \fref{fig:edge-trimming}(a). Using a rotating dicing saw (Disco DAD3113), we removed the mechanically vulnerable peripheral region and cut into the HRSi substrate, limiting the inward propagation of edge damage during thinning. The photographs acquired after donor-substrate removal compare the edge-trimmed and non-edge-trimmed samples (\fref{fig:edge-trimming}(b,c)). Even though both samples revealed the bonding-imperfection-induced edge thin-film vacancy, the edge-trimmed sample retained a larger contiguous transferred area with reduced peripheral delamination.

\section{Simulations for LTOI Mach-Zehnder modulators (MZMs)}
		
\begin{figure}[!htbp]
  \centering
  \includegraphics[width=0.98\linewidth]{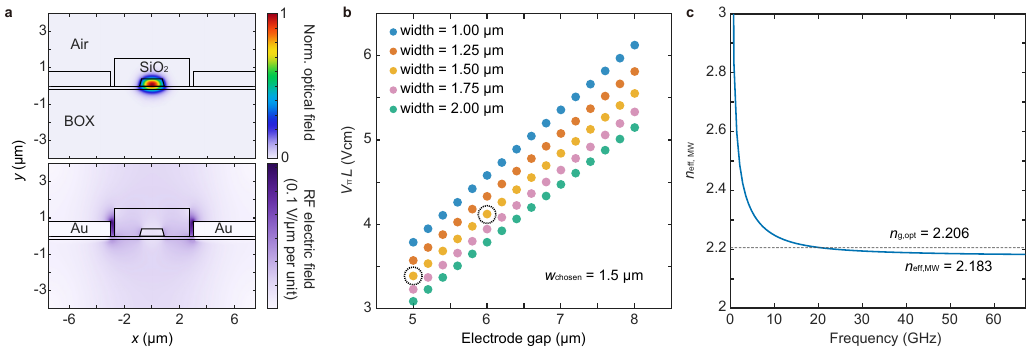}
  \caption{\textbf{Optical and microwave simulations of the traveling-wave LTOI MZM.}
  \textbf{a}, Simulated normalized optical field amplitude, $|E|$, of the fundamental waveguide mode (top) and RF electric-field distribution of the coplanar electrode (bottom). \textbf{b}, Simulated $V_{\pi}L$ as a function of electrode gap for waveguide widths from 1.0~\um{} to 2.0~\um. The dotted circles mark the 1.5-\um-wide devices with 5~\um{} and 6~\um{} electrode gaps used in the measurements. \textbf{c}, Simulated microwave effective index as a function of frequency. The dashed line denotes the optical group index $n_{\mathrm{g,opt}}=2.206$; the high-frequency microwave effective index is $n_{\mathrm{eff,MW}}=2.183$.}
  \label{fig:mzm-simulation}
\end{figure}

The traveling-wave MZM with the push-pull configuration was designed by jointly modeling the fundamental optical mode, the radio-frequency (RF) electric field (\fref{fig:mzm-simulation}(a)) and their phase-matched interaction. The common cross-sectional simulations (COMSOL Multiphysics 6.3) were first employed to calculate its modulation efficiency~\cite{cai2026heterogeneously}. The simulated half-wave-voltage-length product, $V_{\pi}L$, increased with electrode gap over the explored range of 5-8~\um{} (\fref{fig:mzm-simulation}(b)), while increasing the waveguide width from 1.0~\um{} to 2.0~\um{} reduced $V_{\pi}L$ at a fixed electrode gap. The devices characterized here used a 1.5~\um-wide waveguide with either a 5~\um{} or 6~\um{} electrode gap, corresponding to simulated $V_{\pi}L$ values of approximately \SI{3.4}{\volt\centi\meter} and \SI{4.1}{\volt\centi\meter}, respectively.

To maintain the velocity matching behavior across a long propagation distance, the group index of the optical mode must be similar to the effective index of the microwave field~\cite{zhu2021integrated}. Based on a finite-element method (FEM) calculation of a propagating RF field using Ansys HFSS, the microwave effective index ($n_{\mathrm{eff,MW}}$) was extracted from the unwrapped transmission phase of the simulated CPW using the convention $S_{21}\propto\exp(-i\beta L)$:
\begin{equation}
n_{\mathrm{eff,MW}}(f_{\mathrm{RF}})
=-\frac{c_0}{2\pi f_{\mathrm{RF}}L}
\operatorname{unwrap}\!\left[\arg S_{21}(f_{\mathrm{RF}})\right],
\label{eq:mw-index}
\end{equation}
where $c_0$ is the speed of light in vacuum, $f_{\mathrm{RF}}$ is the microwave frequency and $L$ is the length of the CPW electrodes. The minus sign accounts for the decreasing transmission phase with frequency under the stated convention.
At \SI{70}{\giga\hertz}, $n_{\mathrm{eff,MW}}=2.183$, compared with $n_{\mathrm{g,opt}}=2.206$, corresponding to an index difference of 0.023 (\fref{fig:mzm-simulation}(c)). This small mismatch supports broadband velocity matching in the traveling-wave electrode geometry.

\section{Half-wave-voltage characterization}

\begin{figure}[!htbp]
  \centering
  \includegraphics[width=0.96\linewidth]{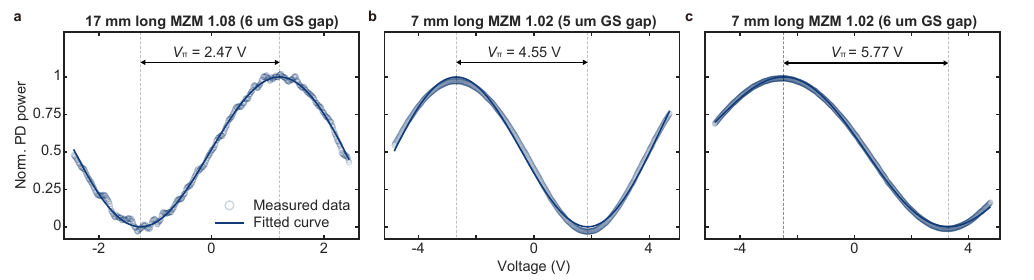}
  \caption{\textbf{Measured half-wave voltages of traveling-wave LTOI MZMs.}
  \textbf{a}, Normalized photodetector power as a function of applied voltage for a \SI{17}{\milli\meter}-long MZM with a 6~\um{} GS gap. \textbf{b,c}, Corresponding transfer curves for \SI{7}{\milli\meter}-long MZMs with 5~\um{} (\textbf{b}) and 6~\um{} (\textbf{c}) GS gaps. Open circles show the measured data, solid curves show sinusoidal fits and arrows mark the fitted half-wave voltages of \SI{2.47}{\volt}, \SI{4.55}{\volt} and \SI{5.77}{\volt}, respectively.}
  \label{fig:vpi}
\end{figure}

For the LTOI MZMs, the linear Pockels effect makes the differential optical phase shift between the two MZM arms proportional to the applied voltage, while interference converts this phase difference into a sinusoidal intensity response. Here, the photodetector power was recorded while the applied bias was swept using a \SI{1}{kHz}, \SI{10}{\volt} peak-to-peak rising sawtooth waveform. The acquired transfer curve was then normalized and fitted to $P(V)=P_{\mathrm{off}}+A[1+\cos(\pi(V-V_0)/V_{\pi})]/2$, with $P_{\mathrm{off}}$, $A$, $V_0$ and $V_{\pi}$ treated as fit parameters. The fitted \(V_{\pi}\), corresponding to a \(\pi\) differential phase shift and equivalently to the voltage separation between adjacent transmission extrema, was taken as the half-wave voltage (\fref{fig:vpi}).

The fits yield $V_{\pi}=\SI{2.47}{\volt}$ for a \SI{17}{\milli\meter}-long device with a 6~\um{} ground-signal (GS) gap, $V_{\pi}=\SI{4.55}{\volt}$ for a \SI{7}{\milli\meter}-long device with a 5~\um{} GS gap and $V_{\pi}=\SI{5.77}{\volt}$ for a \SI{7}{\milli\meter}-long device with a 6~\um{} GS gap. The corresponding $V_{\pi}L$ values are approximately \SI{4.20}{\volt\centi\meter}, \SI{3.19}{\volt\centi\meter} and \SI{4.04}{\volt\centi\meter}, respectively. The similar $V_{\pi}L$ values of the two 6~\um-gap devices are consistent with the expected inverse scaling of $V_{\pi}$ with interaction length. At the same \SI{7}{\milli\meter} length, increasing the GS gap from 5~\um{} to 6~\um{} increased $V_{\pi}$ by approximately 27\%, confirming the reduced electro-optic (EO) field overlap predicted by the above simulation. Overall, the measured \(V_{\pi}L\) values are consistent with the simulated values of approximately \(\SI{3.4}{\volt\centi\meter}\) and \(\SI{4.1}{\volt\centi\meter}\) for the \(5~\um\)- and \(6~\um\)-gap geometries, respectively (\fref{fig:mzm-simulation}(b)).

\section{Modeling of traveling-wave EO responses}

\begin{figure*}[htbp!]
    \centering
    \includegraphics[width=0.98\linewidth]{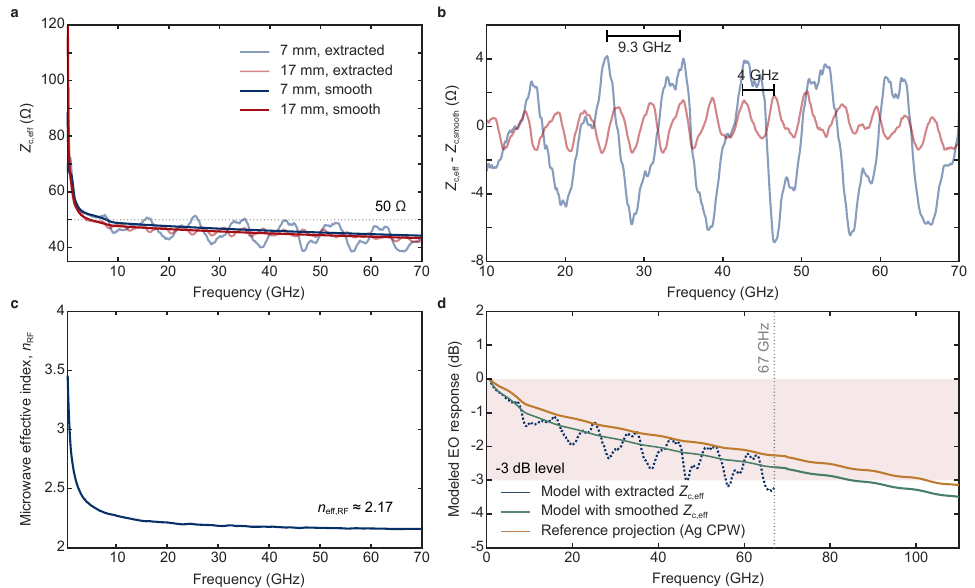}
    \caption{\textbf{Microwave-parameter extraction and EO-response modeling.}
    \textbf{a}, Effective and smoothed characteristic impedance, \(Z_{\mathrm{c,eff}}(f)\), extracted from two-port electrical \(S\)-parameter measurements of the 7- and 17-mm electrodes.
    \textbf{b}, Residual impedance of the 7- and 17-mm electrodes after background subtraction.
    \textbf{c}, Frequency-dependent microwave effective index \(n_{\mathrm{RF}}(f)\) with a representative value of approximately 2.17 between 30 and \(65~\mathrm{GHz}\).
    \textbf{d}, Modeled EO responses calculated using the extracted impedance, smooth-impedance projection and the same projection with the Ag loss profile~\cite{wang2024ultrabroadband}. The two projections yield 3-dB EO bandwidths of \(85.0~\mathrm{GHz}\) and \(102.4~\mathrm{GHz}\), respectively.}
    \label{fig:si-eo-model-analysis}
\end{figure*}

To distinguish the reflection-induced ripple from the overall EO-response roll-off associated with microwave attenuation and velocity mismatch, we extracted the effective characteristic impedance \(Z_{\mathrm{c,eff}}(f)\) and microwave effective index \(n_{\mathrm{RF}}(f)\) from the calibrated EE measurements.
As shown in \fref{fig:si-eo-model-analysis}(a), \(Z_{\mathrm{c,eff}}(f)\) contains a slowly varying background together with pronounced periodic oscillations.
This periodicity was further examined by subtracting the smooth background (\fref{fig:si-eo-model-analysis}(b)), corresponding to approximately \(9.3~\mathrm{GHz}\) and \(4.0~\mathrm{GHz}\) for the 7- and 17-mm electrodes, respectively.
Given the extracted microwave effective index (\fref{fig:si-eo-model-analysis}(c)), \(n_{\mathrm{RF}}\), and electrode round trip, \(\Delta f_{\mathrm{RT}}=c/(2n_{\mathrm{RF}}L)\), the expected frequency period gives approximately \(9.9~\mathrm{GHz}\) and \(4.1~\mathrm{GHz}\), consistent with the extracted residual periods.
This agreement supports coherent reflections between discontinuities separated by approximately the electrode length, identifying the pad-to-CPW transitions as the dominant reflection source.

To fully explore the broadband EO response of our sample beyond the 67 GHz data limit, we employed a standard bidirectional finite-line model~\cite{kharel2021breaking,wang2024ultrabroadband} to calculate the EO transfer function as
\begin{equation}
H_{\mathrm{EO}}(f)=
\frac{1}{L}\int_{0}^{L}
V^{+}(f)
\left[
e^{-\gamma(f)z}
+\Gamma_{\mathrm{L}}(f)e^{-2\gamma(f)L}e^{\gamma(f)z}
\right]
e^{i\beta_{\mathrm{o}}(f)z}\,\mathrm{d}z,
\label{eq:si-eo-finite-line}
\end{equation}
where
\(\gamma(f)=\alpha_{\mathrm{eff}}(f)+i2\pi f n_{\mathrm{RF}}(f)/c\), \(\beta_{\mathrm{o}}(f)=2\pi f n_g/c\), and
\(\Gamma_{\mathrm{L}}(f)=[Z_{\mathrm{L}}-Z_{\mathrm{c}}(f)]/[Z_{\mathrm{L}}+Z_{\mathrm{c}}(f)]\).
Here, \(V^{+}(f)\) is the forward-wave amplitude determined by the standard finite-line boundary conditions with \(Z_{\mathrm{s}}=Z_{\mathrm{L}}=50~\Omega\). The complete extracted \(n_{\mathrm{RF}}(f)\) was retained, while optical group index \(n_g=2.206\) was fixed as obtained from the simulation in \fref{fig:mzm-simulation}(c).
For each device, we fitted one loss parameter to the EO response using the extracted impedance and retained it for the smoothed-impedance projection.
Depending on the calculation, \(Z_{\mathrm{c}}(f)\) denotes either the extracted \(Z_{\mathrm{c,eff}}(f)\) or its smooth background \(Z_{\mathrm{c,smooth}}(f)\).
This finite-line model reproduced both the overall roll-off and the periodic ripple of the measured 7-mm EO response up to 67 GHz (\fref{fig:si-eo-model-analysis}(d)). To assess the response without the transition-related impedance oscillations, only \(Z_{\mathrm{c,eff}}(f)\) was replaced by \(Z_{\mathrm{c,smooth}}(f)\), while \(n_{\mathrm{RF}}(f)\), \(n_g\) and the microwave loss remained unchanged.
The resulting modeled 3-dB EO bandwidth was \(85~\mathrm{GHz}\).
With the electrode geometry unchanged, replacing only the microwave-loss profile with that of the Ag electrode reported in Ref.~\cite{wang2024ultrabroadband} increased the estimated \(3\)-dB EO bandwidth to \(102.4~\mathrm{GHz}\).

\clearpage

\section{Complex microwave-response fitting}

The measured complex reflection coefficient was fitted using the single-port resonator model~\cite{probst2015efficient,bockstiegel2016tunable,kwon2026mechano}
\begin{equation}
S_{11}(\omega)=A \text{e}^{-i\omega\tau}
\left[
1-\frac{\kappa_{\mathrm{ex}}}
{i(\omega-\omega_0)+(\kappa_0+\kappa_{\mathrm{ex}})/2}
\right],
\label{eq:mw-resonator-fit}
\end{equation}
where $A$ is a complex multiplicative background, $\tau$ accounts for the electrical delay of the measurement path, and $\kappa_0$ and $\kappa_{\mathrm{ex}}$ are the intrinsic and external energy-decay rates, respectively. No additive complex background was included. For each device, $\omega_0/2\pi$ was anchored to the measured reflection minimum, and $A$, $\tau$, $\kappa_0$ and $\kappa_{\mathrm{ex}}$ were fitted to the complex $S_{11}$ samples over the full measured frequency range from 57.5 to 63.5 GHz.

Because the coupling regime was not known \textit{a priori}, the complex $S_{11}$ response is fitted to separate $\kappa_0$ from $\kappa_{\mathrm{ex}}$. The resonance linewidth constrains their sum, whereas the background-corrected Smith trajectory distinguishes under- from over-coupling\cite{wang2021cryogenic}.
The quality factors were calculated from the fitted decay rates using $Q_{\mathrm{L}}=\omega_0/(\kappa_0+\kappa_{\mathrm{ex}})$, $Q_0=\omega_0/\kappa_0$ and $Q_{\mathrm{ex}}=\omega_0/\kappa_{\mathrm{ex}}$. For the trap-rich LTOI resonator, the fit yields $Q_{\mathrm{L}}=25.7$, $Q_0=52.5$ and $Q_{\mathrm{ex}}=50.5$; the normal LTOI resonator yielded $Q_{\mathrm{L}}=5.1$, $Q_0=10.1$ and $Q_{\mathrm{ex}}=10.3$.
The corresponding coupling fractions, $\eta=Q_{\mathrm{L}}/Q_{\mathrm{ex}}$, are approximately 0.51 and 0.50, respectively, showing that both devices operate close to critical coupling~\cite{zhang2025ultrabroadband}.

\begin{figure}[!htbp]
  \centering
  \includegraphics[width=0.78\linewidth]{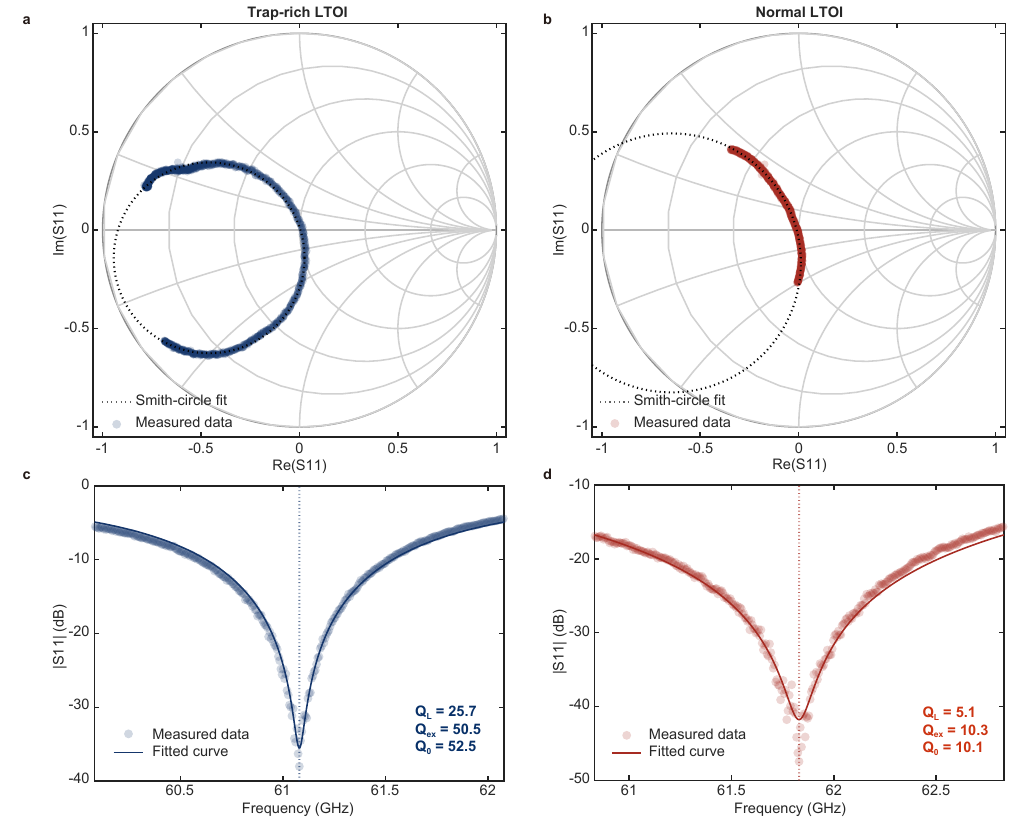}
  \caption{\textbf{Microwave reflection response and resonance fitting.}
  \textbf{a,b}, Complex-plane graphs of the measured reflection coefficients $S_{11}$ for the microwave resonators fabricated on an LTOI wafer with (without) a trap-rich layer \textbf{a} (\textbf{b}). Open circles represent measured data. The dotted circles are independent geometric fits to the measured Smith trajectories and serve only as a visual consistency check. \textbf{c,d}, Corresponding reflection magnitudes and single-port resonator fits for the trap-rich (\textbf{c}) and normal (\textbf{d}) LTOI devices.
  }
  \label{fig:mw-resonance}
\end{figure}

\section{EO-comb-envelope fitting}

The common EO-comb-envelope slope is extracted directly from the raw optical spectrum analyser (OSA) traces (i.e., prior to normalization). For each spectrum, the pump wavelength $\lambda_{\mathrm{p}}$ is identified from the maximum-power sample, and nominal comb-tooth positions are placed on an equally spaced optical-frequency grid using $\lambda_n=c/(\nu_{\mathrm{p}}+n f_{\mathrm{RF}})$, where $\nu_{\mathrm{p}}=c/\lambda_{\mathrm{p}}$. At each position, the maximum raw OSA sample within $\pm45\%$ of the nearest-neighbour spacing was selected without interpolation, smoothing, baseline subtraction or normalization. Teeth above $-60~\mathrm{dBm}$ and within $\SI{5}{\nano\meter}\leq|\lambda_n-\lambda_{\mathrm{p}}|\leq\SI{45}{\nano\meter}$ were retained. All retained teeth were fitted jointly as $P_i^{(\mathrm{dBm})}=b_s+m|\Delta\lambda_i|+\varepsilon_i$, with one common slope $m$ and independent intercepts $b_{\mathrm{L}}$ and $b_{\mathrm{R}}$ for the two wings. The fit used iteratively reweighted least squares with a Huber loss (tuning constant 1.345) and a median-absolute-deviation-based robust scale~\cite{huber1964robust,holland1977robust}. The fitted $m$, expressed in \si{\decibel\per\nano\meter}, was used for the power-dependent comparison in the main text. This model constrains only the two wing decay rates to be equal; it does not impose mirror-symmetric wing amplitudes.

\begin{figure}[!htbp]
  \centering
  \includegraphics[width=0.82\linewidth]{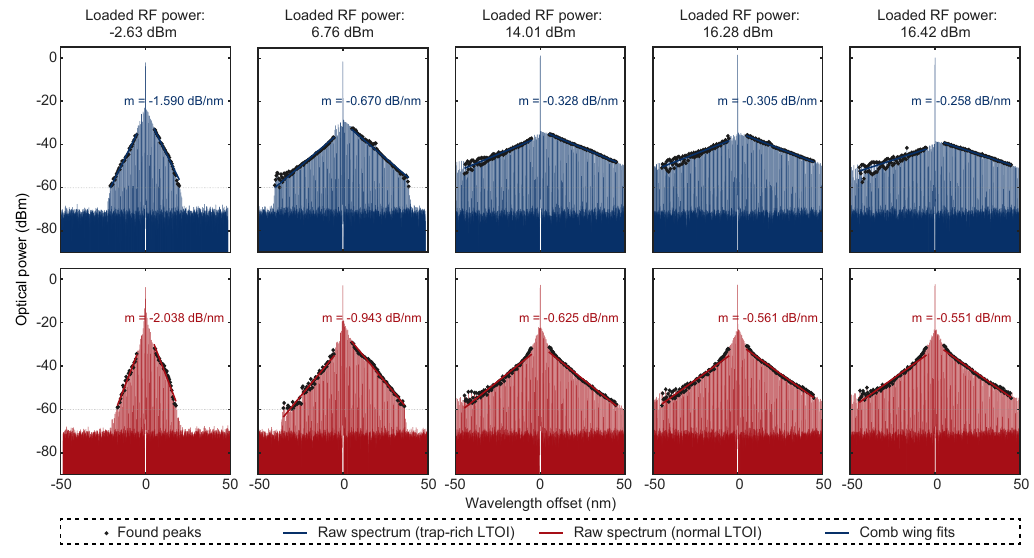}
  \caption{\textbf{Common-slope fitting of EO-comb envelopes.}
  Raw optical spectra and common-slope fits for the trap-rich (top row) and normal (bottom row) LTOI devices at the indicated RF powers. Black markers denote the raw comb teeth selected for fitting. The two solid line segments in each panel share the same slope but have independent intercepts, allowing unequal wing amplitudes without imposing mirror symmetry.}
  \label{fig:comb-envelope}
\end{figure}

\section{EO-response and optical linewidth across the wafer}

\begin{figure}[!htbp]
  \centering
  \includegraphics[width=0.98\linewidth]{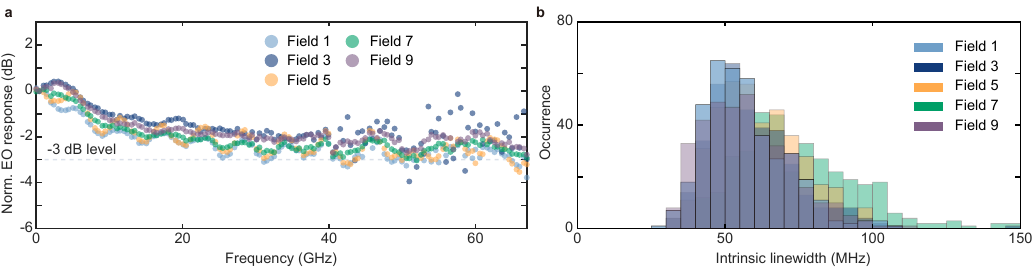}
  \caption{\textbf{Wafer-level EO response and intrinsic-linewidth distributions.}
  \textbf{a}, EO responses of 6~\um-GS-gap MZMs in DUV fields 1, 3, 5, 7 and 9, normalized at \SI{0.5}{\giga\hertz}. \textbf{b}, Intrinsic-linewidth distributions of \SI{100}{\giga\hertz}-free-spectral-range microring resonators in the same fields.}
  \label{fig:wafer-analysis}
\end{figure}

Identical MZMs with a 6~\um{} GS gap and 100-GHz ring resonators in five representative DUV fields (1,3,5,7 and 9) were selected to quantify the device performance uniformity across the wafer.
Figure~\ref{fig:wafer-analysis}(a) displayed the EO responses from five MZMs, which were all normalized to their respective values at $f_0=\SI{0.5}{\giga\hertz}$.
The normalized EO responses followed a similar frequency dependence up to \SI{67}{\giga\hertz}.

Optical-loss uniformity was assessed from the intrinsic-linewidth distributions of \SI{100}{\giga\hertz}-free-spectral-range microring resonators over the optical-frequency range from 185 to 225~THz in the same selected fields (\fref{fig:wafer-analysis}(b)).
The mean linewidths were 57.1~MHz, 60~MHz, 61.2~MHz, 75.6~MHz and 57.7~MHz for Fields 1, 3, 5, 7 and 9, respectively, corresponding to the Q-factor exceeding 2~million. These wafer-level measurements provide the field-resolved distributions underlying the representative EO response and optical linewidth results as reported in the main text.

\section{Cryogenic T-resonator design and fitting}

To probe substrate-associated microwave loss over temperature, the resonator geometry had to provide a measurable response for devices with unknown internal losses.
In this work, the large loss contrast between the trap-rich and normal LTOI substrates makes it difficult to choose a single external decay rate, $\kappa_{\mathrm{ex}}$, comparable to both internal rates, $\kappa_0$.
Tunable couplers avoid this mismatch but require additional circuitry and control~\cite{bockstiegel2016tunable}.
We therefore used a directly connected coplanar-waveguide T resonator, which removes the separate weak-coupling element and maps the distributed stub loss onto the complex through response~\cite{peterson2001cpw}.

\begin{figure}[!htbp]
  \centering
  \includegraphics[width=0.98\linewidth]{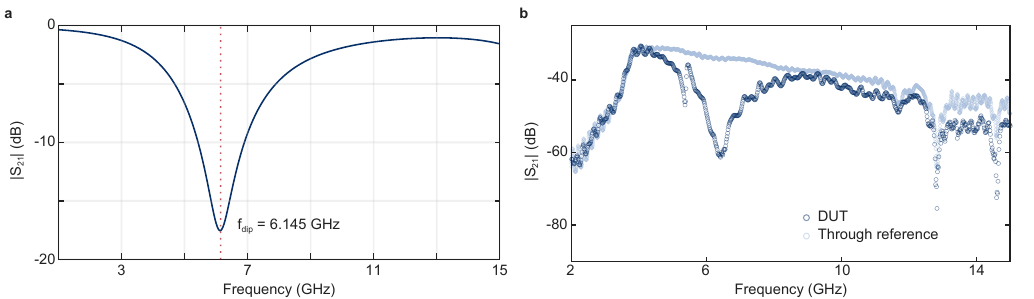}
  \caption{\textbf{T-resonator simulation and calibration.}
  \textbf{a}, Simulated transmission magnitude, showing a broad notch at \SI{6.145}{\giga\hertz}. \textbf{b}, Representative device-under-test (DUT) and device-free through responses used for complex de-embedding of the cryogenic microwave link.}
  \label{fig:t-resonator}
\end{figure}

The T resonator consists of a 5-mm-long open-ended quarter-wave stub connected in shunt to a nominally \(50~\Omega\) through line, with the stub treated as a uniform quasi-TEM transmission line and the through line assumed to be impedance matched to the input and output reference ports.
The stub input admittance is \(Y_{\mathrm{in}}=Y_c\tanh(\gamma L)\), where \(Y_c\) is the characteristic admittance of the stub, \(L\) is its physical length, and \(\gamma=\alpha_{\mathrm{eff}}+i\beta\) is the complex propagation constant, with \(\alpha_{\mathrm{eff}}\) denoting the effective amplitude attenuation per unit length and \(\beta\) denoting the phase constant.
At the fundamental quarter-wave resonance, \(\beta L\approx\pi/2\), the open termination is transformed into an effective short circuit at the T junction, producing a large shunt admittance and a transmission notch.
Assuming an approximately frequency-independent phase velocity, \(\beta L\approx\pi f/(2f_r)\), the complex forward transmission is expressed as
\begin{equation}
S_{21}(f)=B(f)\frac{2}{2+\kappa\tanh\!\left(\alpha_{\mathrm{eff}}L+i\frac{\pi f}{2f_r}\right)},
\label{eq:stub-response}
\end{equation}
where $f_r$ is the effective electrical quarter-wave resonance frequency, and $\kappa$ is the effective junction-coupling factor distinct from \(\kappa_{\mathrm{ex}}\). The slowly varying complex prefactor \(B(f)\) accounts for the residual broadband response of the measurement chain after calibration and is not assigned a physical resonator-loss interpretation.
As shown in \fref{fig:t-resonator}(a), full-wave electromagnetic simulations (Ansys HFSS) predict a broad transmission notch near 6.145 GHz.
Given the several-gigahertz span of the notch, the broadband calibration of the cryogenic microwave link is essential for reliable parameter extraction.
The complex DUT trace was therefore de-embedded using the device-free through trace shown in \fref{fig:t-resonator}(b) and normalized to 0 dB. The resulting complex response was fitted directly to the distributed T-resonator model described in Eq.~\ref{eq:stub-response}, with the stub length \(L\) fixed to its design value and \(\kappa\), \(\alpha_{\mathrm{eff}}\), and \(f_r\) as the fitted parameters.

\clearpage

\bigskip
\hypersetup{urlcolor=black}
\bibliographystyle{apsrev4-2}
\bibliography{references}